\documentclass{mem}
\usepackage{natbib}\usepackage{txfonts}\usepackage{balance}
\usepackage{graphicx}
\usepackage[a4paper,breaklinks,dvipdfm]{hyperref}

\idline{75}{1}
\begin{document}
\def\teff{$T\rm_{eff }$}
\def\kms{$\mathrm {km s}^{-1}$}

\title{
Optical and X-ray Variability of the Peculiar Cataclysmic Variable FS Aur with
a Magnetic and Freely Precessing White Dwarf}

\subtitle{}

\author{
V.\,Neustroev\inst{1},
G.\,Tovmassian\inst{2},
S.\,Zharikov\inst{2},
G.\,Sjoberg\inst{3,4},
T.\,Arranz Heras\inst{5},
P.\,B.\,Lake\inst{4},
D.\,Lane\inst{6,7},
G.\,Lubcke\inst{4},
\and A.\.A.\,Henden\inst{4}
}

\offprints{V. Neustroev, \email{vitaly@neustroev.net}}

\institute{
Astronomy Div., Dept. of Physics, PO Box 3000, FIN-90014 University of Oulu,
Finland
\and
Inst. de Astronomia, UNAM, Apdo. Postal 877, Ensenada, Baja California, 22800 Mexico
\and
The George-Elma Observatory, Mayhill, New Mexico, USA
\and
AAVSO, 49 Bay State Road, Cambridge, MA 02138, USA
\and
Observatorio ``Las Pegueras'', Navas de Oro (Segovia), Spain
\and
Saint Mary's University, Halifax, Nova Scotia, Canada
\and
The Abbey Ridge Observatory, Stillwater Lake, Nova Scotia, Canada
}

\authorrunning{Neustroev et al.}

\titlerunning{Optical and X-ray variability of FS Aur}

\abstract{
We present preliminary results of long-term monitoring of the peculiar
cataclysmic variable FS Aurigae conducted during the 2010-2011 observational
season. The multicolor observations revealed, for the first
time in photometric data, the precession period of the white dwarf,
previously seen only spectroscopically.
This is best seen in the $B-I$ color index and reflects the
spectral energy distribution variability. Analysis of X-ray
observations made with Chandra and Swift, also revealed the
existence of both the orbital and precession periods in the light curve.
We also show that
the long-term variability of FS Aur and the character of its outburst
activity may be caused by variations in the mass transfer rate from
the secondary star as the result of eccentricity modulation of a close
binary orbit induced by the presence of a third body on a circumbinary
orbit.

}
\maketitle{}

\begin{figure*}
\begin{center}
\includegraphics[width=1.0\textwidth]{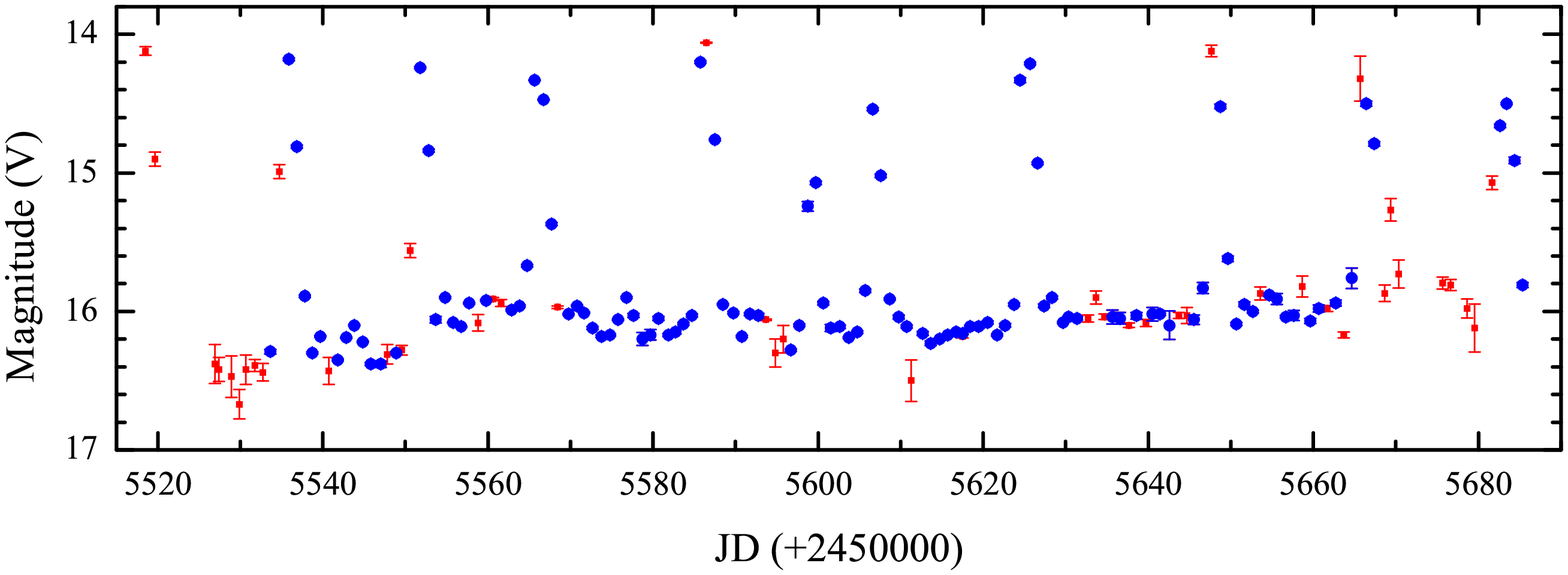}
\caption{Light curve of FS Aur, from the 2010-2011 observing campaign.
Each point is the 1-day average of observations. Blue points represent
the observations obtained by the authors, while the red squares represent
the AAVSO observations.}
\label{fig:1daver}
\end{center}
\end{figure*}

\section{Introduction}

FS Aurigae represents one of the most unusual cataclysmic variable (CV)
to have ever been observed. The system is famous for a variety of uncommon
and puzzling periodic photometric and spectroscopic variabilities which do
not fit well into any of the established sub-classes of CVs.

The outlandish peculiarity of FS Aur is the existence of well-defined
photometric optical modulations with the amplitude of up to $\sim0.5$ mag and
a very coherent long photometric period (LPP) of 205.5 min that exceeds the
spectroscopic orbital period (OP) of 85.7 min by 2.4 times
\citep{Neustroev2002,Tovmassian2003}. Such a disagreement in the photometric
and spectroscopic periods is highly unusual for a low mass binary system that
is unambiguously identified as a CV.

During the spectral observations made in December 2004, \citet{Tovmassian2007}
discovered a second long spectroscopic period (LSP) of 147 minutes, appearing
in the far wings of the emission lines. They showed that frequency of this new
period is equal exactly to the beat between the OP and LPP:
$1/P_{beat}=1/P_{orb} - 1/P_{phot}$.

\citet{Tovmassian2007} proposed that an interpretation of this puzzling
behaviour of FS~Aur might be a rapidly rotating magnetic WD precessing with the
LSP. The spectroscopic period is a direct measure of
rotation of the inner bright spot (precession period according to our
hypothesis) and that the period seen in photometry (LPP) is the beat.
According to existing models \citep{Leins}, the WD rotational period should be 
on the order of 50 - 100 sec in order to have the proposed precession period.
\citet{Neustroev2005} and \citet{Tovmassian2010,Tovmassian2012} intended
to reveal the spin period of the WD by fast optical and X-ray photometry. They
found a $\sim$101 sec peak in both the power spectra, but it was not conclusive
evidence.

Based on the short orbital period, FS~Aur has been classified as a SU UMa star.
Nevertheless, long-term monitoring of the system failed to detect any
superoutburst/superhumps in its light curve. Instead, this monitoring revealed
a very long photometric period of $\sim$900 days. \citet{Tovmassian2010} showed
that such a long period may be explained by the presence of sub-stellar third
body on a circular orbit around the close binary.

In order to better understand the long-term variability and outburst activity of FS~Aur,
during the 2010-2011 observational season we have initiated and conducted an observing
campaign, lasting more than 140 consecutive nights. Here we report preliminary results
of these observations.

\section{The 2010-2011 observing campaign}

The observations were conducted every clear
night from November 26, 2010 until May 3, 2011. The data were taken using telescopes
with apertures of 0.28 to 0.5-meters, equipped with CCD cameras and standard Johnson
V filters. Depending on the weather conditions, we monitored the star for 6--8 hours
per night in the beginning of the campaign and for 3--4 hours at the end. Thus, more
than 80 nights of time-resolved photometry were taken, and almost 14\,000 V-band data
points were obtained. In order to reduce the scatter from both random errors and
stochastic and short-term variability, we formed 1-day averages of these observations
(Fig.~\ref{fig:1daver}).

Additionally, between nights of January 20 and March 9, 2011, 31 nights of time-resolved
multicolor $B, V, R_c, I_c$ photometry were taken. Moreover, after acceptance of
our Swift ToO request, we obtained 5 more nights of multicolor observations, between
March 28 and April 3, 2011.

\begin{figure}
\centerline{\includegraphics[width=6.5cm]{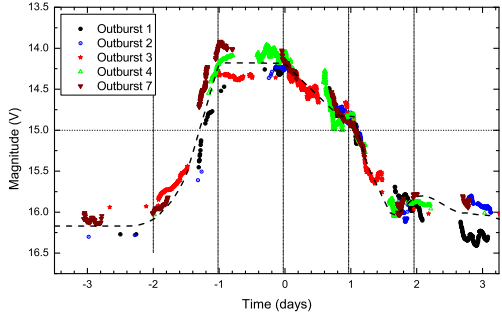}}
\caption{The mean outburst profile of FS Aur is an average of 5 best covered outbursts.}
\label{fig:outbursts}
\end{figure}

\begin{figure}[h]
\begin{center}
\hbox{
\includegraphics[width=6.5cm]{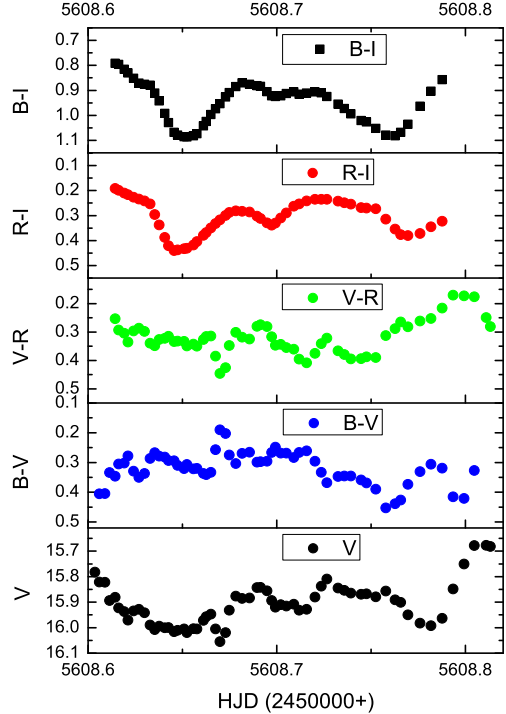}
}
\end{center}
\caption{\footnotesize
{A sample $V$ light curve and color curves $B-V$, $V-R_c$, $R_c-I_c$ and
$B-I_c$ from JD 2\,455\,608.}}
\label{fig:colors}
\end{figure}

\begin{figure}
\centerline{\includegraphics[width=6.5cm]{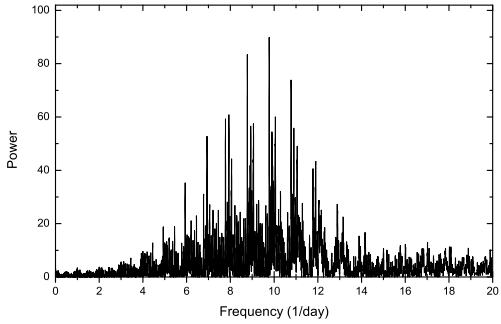}}
\caption{The Lomb-Scargle power spectrum of the $B-I$ color-index curve of FS Aur.
The strongest peak at $f$ = 9.776 $\pm$ 0.003 day$^{-1}$ exactly coincides with
the LSP which is the presumed precession period of the WD.}
\label{scargle}
\end{figure}

\section{Long-term variability and outburst activity of FS Aurigae}

During the campaign, we succeeded in observing 11 consecutive outbursts, most of
them in great details. Here we note the most prominent features seen in both
the outbursts and quiescent light curves:

\begin{enumerate}
\item All the observed outbursts were of low amplitude ($\leq$ 2 mag) that is lower
than most dwarf novae exhibit;
\item Nine of eleven outbursts had a very similar shape to the light curve:
the fast rise (2.7 mag d$^{-1}$), a flat maximum, the relatively slow decline
(0.8 mag d$^{-1}$), the rate of which, after the short plateau phase, dramatically
increased to $\sim$2.0 mag d$^{-1}$. The duration of each of these
stages is around one day (Fig~\ref{fig:outbursts});
\item Two outbursts were of abnormally low amplitude and of short duration
(around JD\,2455600, Fig.~\ref{fig:1daver});
\item The recurrence time of the normal outbursts was relatively short and stable,
$18.1\pm2.5$ d;
\item During the decline stage of all the observed outbursts, large-amplitude
modulations of 0.25--0.35 mag with a recurrence period of some 10 per cent
longer than the OP (``superhumps'') appeared in the light curve. They were
observed for the most part of the quiescent stage but, approaching a next
outburst, the superhumps usually disappeared to be replaced by
the LPP modulations;
\item The system showed strong short-term variability during all stages
of outbursts and quiescence;
\item During these observations, the average quiescent level increased
$0.3-0.4$ mag.
\end{enumerate}

The strong variability during the quiescent state may be due to the variable
mass transfer rate. The optical flux of a dwarf nova is dominated by the
emission from an accretion disk particularly in the short period systems,
while the emission from the disk is proportional to the mass-transfer rate.
\citet{Schreiber2000} also found that the outburst behavior of a dwarf nova
is strongly influenced by the variations of the mass-transfer rate.
The latter is very sensitive to the Roche lobe size, which is proportional to
the binary separation. In hierarchical triple systems,
a third body can induce an eccentricity variation in an inner binary
\citep{Mazeh,Georgakarakos2009}. The long-term modulation is produced by the
time-varying tidal force of the perturber upon the binary.

\section{Optical and X-ray variability with the precession period}

Usually, the most prominent features of the optical light curve of FS Aur is
the well-defined LPP modulations of 205.5 min contaminated by strong stochastic
variations, while the orbital variability is seen only occasionally
\citep{Tovmassian2003, Neustroev2005, Neustroev2011}.
However, during the presented observations the photometric behavior of FS Aur was
even more complex than usually. In particular, during the most part of the quiescent
stage, we observed modulations with  a period of some 10 per cent longer than the OP.

This brightness variation was approximately the same in all filters. However,
the color curves $B-V$, $V-R_c$ and $R_c-I_c$ also show a significant variation,
more or less following the brightness variation: the color indexes $B-V$ and
$R_c-I_c$ are generally anti-correlated and ${V-R_c}$ is correlated with the light
curve (Fig.~\ref{fig:colors}). The amplitudes of color variations are
$\sim0.05-0.1$ in all these colors.

Nevertheless, the behavior of the color index $B-I_c$ is significantly different.
The most noticeable feature in $B-I_c$ is a modulation with a period near 0.1 day,
even though it is sometimes contaminated or even completely replaced by the
orbital/superhump variations. After visual inspection of the nightly $B-I_c$
curves, we selected a total of 16 nights (two-third of the observations taken in
quiescence) when this variability is most evident. The Lomb-Scargle periodogram
of the combined data (with the means, and linear trends subtracted for each night)
is shown in Fig.~\ref{scargle}.

The strongest peak at $f = 9.776 \pm$ 0.003 day$^{-1}$
exactly coincides with the beat period between the OP and LPP: $1/P_{beat}=1/P_{orb} -
1/P_{phot}$, which is the presumed precession period of the WD in FS~Aur.
Variability with such a period was previously observed only in spectroscopic data
as the Long Spectroscopic Period \citep{Tovmassian2007}.

During the PP, the properties of the X-ray and UV spectrum are expected
to be even more variable because the accretion column on the WD magnetic
pole is observed from different viewing angles.

Motivated by our findings, we performed a Swift ToO observation of FS~Aur
during 2011 March 29 -- April 1 for a total of about 20.4 ksec. The object
was also observed by Chandra in 2005 (25 ksec), and by Swift in 2007 (30 ksec).
FS Aur is a relatively bright X-ray source with a count-rate of 0.38 cnts/s as
observed with Chandra's ACIS-S3 CCD array. The important result of the X-ray
observations is the detection of the modulations with the OP. Those X-ray 
modulations have a pulse profile and are stronger in the soft band
(see Figs~3 and 4 in \citealt{Tovmassian2012}). We note that one out of
five pulses observed by Chandra, seems to be missing in the light curve.

\begin{figure}
\label{fig:xray}
\begin{center}
\hbox{
\includegraphics[width=6.5cm]{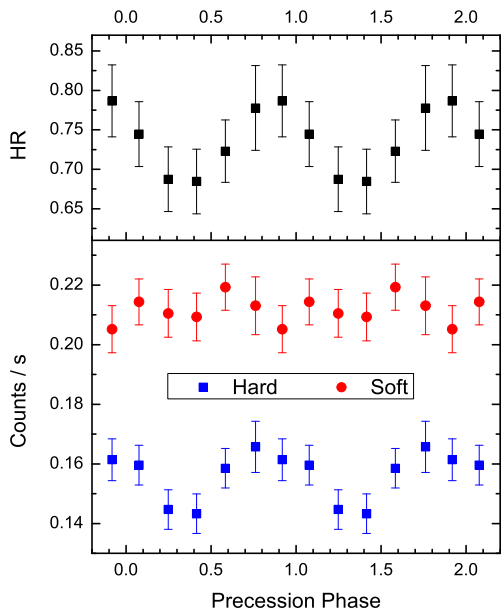}
}
\end{center}
\caption{\footnotesize
{
The Chandra X-ray light curve and hardness ratio curve of FS Aurigae,
folded with the precession period.}
}
\end{figure}

In order to look for variability with the PP, we folded the X-ray data
with the period of 147.3 min. There is a strong signature of modulation,
mostly seen in the hard band (Fig.~\ref{fig:xray}). What is remarkable
is that the omission of one of the pulses in the Chandra X-ray light curve
occurred around the time of the flux minimum during the precession period
(Fig.~3 in \citealt{Tovmassian2012}).

\section{Conclusion}

Our new multicolor observations made during the 2010-2011 observational season
revealed, for the first time in photometric data, the precession period of the
WD, previously seen only spectroscopically. This is best seen in the $B-I$ color
index and reflects the spectral energy distribution variability. Analysis of
X-ray observations made with Chandra and Swift, also revealed the existence of
both the orbital and precession periods.

\citet{Tovmassian2010} showed that a long $\sim$900-d period observed in FS Aur
may be explained by the presence of a sub-stellar third body on a circular orbit
around the close binary. The long-term variability of FS Aur and the character
of its outburst activity may also be triggered by variations in $\dot{M}$ from
the secondary as the result of eccentricity modulation of a close binary orbit
induced by the presence of a third body.

\begin{acknowledgements}
We thank Neil Gehrels for approving the Target of Opportunity observation with
Swift and the Swift team for executing the observation.
We acknowledge with thanks the variable star observations from the AAVSO
International Database contributed by observers worldwide and used in this research.
\end{acknowledgements}

\bibliographystyle{aa}

\end{document}